\documentclass
[%
reprint,
 amsmath,amssymb,
 aps,
]{revtex4-1}

\usepackage{graphicx}
\usepackage{dcolumn}
\usepackage{bm}
\usepackage{hyperref}
\usepackage{csquotes}
\usepackage{xcolor}
\usepackage{pgfplots}
\usepackage[normalem]{ulem}

\begin{document}

\title{Dynamical Resolution of the Cosmic Coincidence Problem in Non-Interacting Holographic Dark Energy via Einstein–Cartan Torsion}

\author{Yongjun Yun}
\affiliation{Graduate School Department of Physics, Daejin University, Pocheon 11159, Korea}

\author{Kyungduk Kim}
\affiliation{School of Liberal Arts and Sciences, Korea National University of Transportation, Chungju 27469, Korea}

\author{Jungjai Lee}
\email{jjlee@daejin.ac.kr}
\affiliation{Graduate School Department of Physics, Daejin University, Pocheon 11159, Korea}
\affiliation{School of Physics, Korea Institute for Advanced Study, Seoul 02455, Korea}

\begin{abstract}
We investigate the cosmic coincidence problem in non-interacting holographic dark energy with the Hubble radius as the infrared cutoff in Einstein-Cartan gravity. In general relativity, this cutoff gives a dust-like equation of state in the non-interacting case, whereas interacting models require a phenomenological dark sector coupling and yield a constant density ratio. We show that the Einstein-Cartan torsion scalar $\Phi$, compatible with the cosmological principle, with the self-consistent scaling behavior $\Phi\sim a^{-3}$, makes the density ratio $r\equiv \rho_m/\rho_X$ dynamical even when the phenomenological interaction term is absent, $Q=0$. The same torsion contribution shifts the equation of state for holographic dark energy toward negative values, allowing cosmic
acceleration, and realizes the observed order-unity density ratio within the weak torsion regime without tuning the holographic free parameter. Thus, Einstein-Cartan torsion provides a geometric mechanism that replaces the phenomenological dark sector interaction and offers a dynamical resolution of the cosmic coincidence problem.
\end{abstract}

\maketitle

\section{Introduction}
The observed accelerated expansion of the universe \cite{1,2} is commonly attributed to a dark energy component with negative pressure. The simplest and most conventional realization of such a component is the cosmological constant $\Lambda$. 
Despite its phenomenological success, however, the cosmological constant model faces two well-known conceptual difficulties, namely the fine-tuning problem \cite{3} and the cosmic coincidence problem \cite{4}. 
The latter asks why the matter density and the dark energy density are of the same order at the current epoch. 
Holographic dark energy provides one possible framework for addressing these problems \cite{6}.

Motivated by the holographic principle \cite{7}, holographic dark energy relates the vacuum energy density to an infrared (IR) length scale.
When the future event horizon is adopted as the IR cutoff, non-interacting holographic dark energy models can successfully drive cosmic acceleration \cite{8}, but they are known to suffer from the causality problem and circular logic problem \cite{9}.
Nevertheless, the cosmic coincidence problem has also been addressed within this framework by linking the onset of late-time acceleration to early inflation \cite{10}.

This motivates the consideration of alternative IR cutoffs. 
Among them, the Hubble radius is a particularly simple choice because it is determined entirely by the local expansion rate of the universe.
In non-interacting holographic dark energy models with the Hubble radius as the IR cutoff, the matter density and the dark energy density exhibit the same scaling behavior, so that the equation of state for dark energy coincides with that of matter, $\omega_X=\omega_m$ \cite{6,11}. 
Therefore, the model cannot drive cosmic acceleration.

Introducing an interaction between dark matter and dark energy can overcome this difficulty. 
In terms of the decay rate $\Gamma$ associated with the
interaction between the dark sectors, the equation of state is modified as \cite{12}
\begin{equation} \label{e.o.s Gamma}
    \omega_X
    = \omega_m - \left( \frac{r+1}{r} \right) \frac{\Gamma}{3H},
\end{equation}
where $H$ is the Hubble parameter and $r$ is the density ratio of matter density to dark energy density.
For $\Gamma>0$, the equation of state is shifted toward negative values, allowing for the possibility of cosmic acceleration.
However, the interacting model suffers from two limitations, namely the absence of a non-interacting limit and a constant density ratio, $r=\mathrm{constant}$ \cite{12}.
The latter prevents a dynamical resolution of the cosmic coincidence problem.
Moreover, interacting models may face additional constraints when one attempts to address the Hubble tension and the $S_8$ tension simultaneously \cite{13,14}, which further motivates a strictly non-interacting mechanism.

Although there is no well-established theory of quantum gravity today, our goal is to analyze the dark energy problem and the cosmic coincidence problem using a model that is as close as possible to quantum gravity. 
Notably, the classical limit of quantum gravity may pass through Einstein-Cartan theory before reducing to general relativity, and this possibility offers a different route to non-interacting holographic dark energy.

In Einstein-Cartan theory, because torsion is algebraically related to the intrinsic spin of matter, it can modify the cosmological dynamics without introducing a phenomenological interaction between dark sectors.
In a recent work \cite{15}, it was shown that a torsion scalar compatible with the cosmological principle allows non-interacting holographic dark energy with the Hubble radius as the IR cutoff to drive cosmic acceleration. 
In that framework, a scaling relation of the torsion scalar $\Phi$ with the scale factor $a$, $\Phi \sim a^{-3}$, has been derived self-consistently without imposing an ad hoc ansatz.

In this Letter, we investigate whether non-interacting holographic dark energy with the Hubble radius as the IR cutoff can provide a dynamical resolution of the cosmic coincidence problem in the presence of the torsion scalar.
We show that the torsion scalar contribution makes the density ratio between matter and dark energy time-dependent even in the absence of interaction between dark matter and dark energy, while also enabling cosmic acceleration.
As a result, the order-unity value of the current density ratio can be interpreted as a consequence of torsion-induced dynamical evolution, rather than as a fixed consequence of the free parameter.

In Sec. II, we examine how the torsion scalar affects the density ratio of the matter density to the dark energy density, even in the absence of interaction between the dark sectors, and discuss a resolution to the cosmic coincidence problem.

\section{Holographic Dark Energy Model in Einstein-Cartan Theory}
In the Einstein-Cartan theory, the torsion tensor is defined as $S_{\mu\nu}{}^\rho = \Gamma^\rho_{[\mu\nu]}$, where $\Gamma^\rho_{\mu\nu}$ is the affine connection.
The torsion tensor that preserves the cosmological principle is given by (see \cite{16,17} for more details)
\begin{equation} \label{torsion scalar}
    S_{\mu\nu\rho}
    = \Phi(t) h_{\rho[\mu} u_{\nu]},
\end{equation}
where $\Phi(t)$ is a torsion scalar \cite{15}, $u^\mu$ is a four-velocity vector, and $h_{\mu\nu}$ is the projection tensor orthogonal to $u^\mu$.
Under the cosmological principle, we can use the flat FLRW metric
\begin{equation} \label{FLRW}
	ds^2
	= - dt^2
	+ a^2(t) \left( dr^2 + r^2 d\theta^2 + r^2 \sin^2 \theta d\varphi^2 \right),
\end{equation}
where $a$ is the scale factor.

The Einstein-Cartan field equations and the Cartan equations are obtained from the action principle. The latter are algebraic equations relating the torsion tensor to the spin tensor of matter, so that torsion is sourced by intrinsic spin rather than by a phenomenological dark-sector interaction.
For a Weyssenhoff spin fluid, the resulting spin-torsion structure gives the Friedmann-like equations \cite{15}
\begin{equation} \label{Friedmann-like 1}
	H^2
	= \frac{1}{3 M_p^2} \left( \rho - 3 M_p^2 \Phi^2 \right)
\end{equation}
and
\begin{equation} \label{Friedmann-like 2}
	\dot{H} + H^2
	= - \frac{1}{6 M_p^2} \left( \rho + 3 p - 12 M_p^2 \Phi^2 \right),
\end{equation}
where $H =\dot{a}/a$ is the Hubble parameter and $M_p = 1 / \sqrt{8 \pi G}$ is the reduced Planck mass.
The total energy density and pressure are expressed as $\rho = \rho_m + \rho_X$ and $p = p_m + p_X$, where $m$ and $X$ denote matter and dark energy, respectively.
With the critical density $\rho_c = 3 M_p^2 H^2$, the density parameters are defined as $\Omega_m = \rho_m / \rho_c$ and $\Omega_X = \rho_X / \rho_c$.
Then, the first Friedmann-like equation can be recast as
\begin{equation}
    \Omega_m + \Omega_X - \left(\frac{\Phi}{H}\right)^{2} = 1.
\end{equation}
Within this framework, the evolution equation for the torsion scalar is obtained self-consistently as
\begin{equation} \label{evolution torsion scalar}
    \dot{\Phi} + 3 H \Phi = 0,
\end{equation}
which gives the scaling behavior \cite{15}, namely
\begin{equation} \label{torsion scalar scaling behavior}
    \Phi \sim a^{-3}.
\end{equation}
It is important to note that this scaling behavior is determined without introducing an ad hoc ansatz.
Combining the two Friedmann-like equations \eqref{Friedmann-like 1} and \eqref{Friedmann-like 2} with Eq. \eqref{evolution torsion scalar} yields the continuity equation
\begin{equation}
    \dot{\rho} + 3 H \left( \rho + p \right)
    = 0,
\end{equation}
which can be decomposed into
\begin{equation} \label{continuity eq of matter}
	\dot{\rho}_{m} + 3 H \left( \rho_m + p_m \right) 
    = Q
\end{equation}
and
\begin{equation} \label{continuity eq for dark energy}
	\dot{\rho}_{X} + 3 H \left( \rho_X + p_X \right)
    = - Q,
\end{equation}
where $Q$ is a phenomenological interaction term.
Since a fundamental theory of dark matter and dark energy has not yet been established, the interaction term cannot be derived from the first principle \cite{6}.
It is introduced phenomenologically, but such interactions still lack observational evidence.
Moreover, models with such interactions may face conflicting constraints when attempting to resolve the Hubble tension and $S_8$ tension simultaneously \cite{13,14}.
For these reasons, we focus on the non-interacting case, namely
\begin{equation} \label{continuity eq of matter no inter}
	\dot{\rho}_{m} + 3 H \rho_{m} \left( 1 + \omega_{m} \right) 
    = 0
\end{equation}
and
\begin{equation} \label{continuity eq for dark energy no inter}
	\dot{\rho}_{X} + 3 H \rho_{X} \left( 1 + \omega_{X} \right) 
    = 0,
\end{equation}
where $\omega_m = p_m / \rho_m$ and $\omega_X = p_X / \rho_X$ are the equations of state, respectively.

Cohen et al. \cite{7} suggested that vacuum energy in a region of finite size $L$ should not exceed the mass of a black hole of the same size, $L^{3}\rho_X \lesssim LM_{p}^{2}$.
Saturating this bound produces a holographic dark energy $\rho_X = 3d^2M_p^2L^{-2}$ \cite{8}, where $d$ is a free parameter.
Setting the IR cutoff to the Hubble radius, $L=H^{-1}$, leads to
\begin{equation} \label{holographic DE}
    \rho_X = 3d^2M_p^2H^2.
\end{equation} 
The density parameter for the dark energy is given by $\Omega_X = \rho_X / \rho_c = d^2$.
Then, the first Friedmann-like equation \eqref{Friedmann-like 1} reads 
\begin{equation}
    \Omega_m + d^2 - 1 = \left(\frac{\Phi}{H}\right)^{2}.
\end{equation}
Since $(\Phi/H)^{2} > 0$, one finds
\begin{equation}
    d^2 > 1 - \Omega_m.
\end{equation}
For the observed value $\Omega_m^0 \simeq 0.3$ \cite{18}, the free parameter is restricted to $0.837 \lesssim d$.
Since previous estimates suggest that the free parameter $d$ is close to unity \cite{19,20}, while values $d>1$ are disfavored by the second law of thermodynamics \cite{8}, we restrict our analysis to
$0.837 \lesssim d < 1$ in this work.

Using the Friedmann-like equation \eqref{Friedmann-like 1} together with the holographic constraint \eqref{holographic DE}, the matter density takes the form
\begin{equation}
    \rho_{m} 
    = 3M_{p}^{2}H^{2}\left[1-d^{2}+\left(\frac{\Phi}{H}\right)^{2}\right],
\end{equation}
and thus the density ratio of the matter density to the dark
energy density is given by
\begin{equation} \label{ratio torsion}
    r
    = \frac{\rho_m}{\rho_X}
    = \frac{1-d^2+\left(\frac{\Phi}{H}\right)^2}{d^2}.
\end{equation}
In the presence of the torsion scalar, the density ratio is no longer constant and becomes time-dependent, even in the absence of interaction between dark matter and dark energy.
This opens up the possibility of a dynamical resolution of the cosmic coincidence problem.
Differentiating Eq. \eqref{ratio torsion} together with Eq. \eqref{evolution torsion scalar}, one obtains
\begin{equation} \label{dot r 1}
    \frac{\dot{r}}{r} 
    = - 2H\left(2-q\right) \frac{\left(\frac{\Phi}{H}\right)^2}{1-d^2+\left(\frac{\Phi}{H}\right)^2},
\end{equation}
where $q = - \ddot{a}a / \dot{a}^2 = - (\dot{H}+H^2)/H^2$ is the deceleration parameter.
Using the two continuity equations \eqref{continuity eq of matter no inter} and \eqref{continuity eq for dark energy no inter}, the evolution of the density ratio can be expressed as
\begin{equation} \label{dot r 2}
    \frac{\dot{r}}{r}
    = \frac{\dot{\rho}_m}{\rho_m} 
    - \frac{\dot{\rho}_X}{\rho_X}
    = 3 H \left( \omega_X - \omega_m \right).
\end{equation}
Combining Eqs. \eqref{dot r 1} and \eqref{dot r 2} yields the equation of state
\begin{equation} \label{e.o.s}
    \omega_{X} 
    = \omega_{m} 
    - \frac{2}{3}\left(2-q\right)\frac{\left(\frac{\Phi}{H}\right)^2}{1-d^2+\left(\frac{\Phi}{H}\right)^2},
\end{equation}
which is equivalent to
\begin{equation} \label{e.o.s torsion scalar}
    \omega_{X} 
    = \omega_{m} - \left(\frac{r+1}{r}\right) \frac{1}{3H} \left[2H(2-q)\frac{\left(\frac{\Phi}{H}\right)^2}{1+\left(\frac{\Phi}{H}\right)^2}\right].
\end{equation}
Comparison of Eq. \eqref{e.o.s torsion scalar} with Eq. \eqref{e.o.s Gamma} shows that the term containing the torsion scalar in square bracket of Eq. \eqref{e.o.s torsion scalar} plays a role analogous to the decay rate.
This indicates that some effects of the interaction between dark matter and dark energy in general relativity can be mimicked by the torsion scalar in Einstein-Cartan theory without introducing such an interaction.

\begin{figure}[h]
    \centering
    \includegraphics[width=0.8\linewidth]{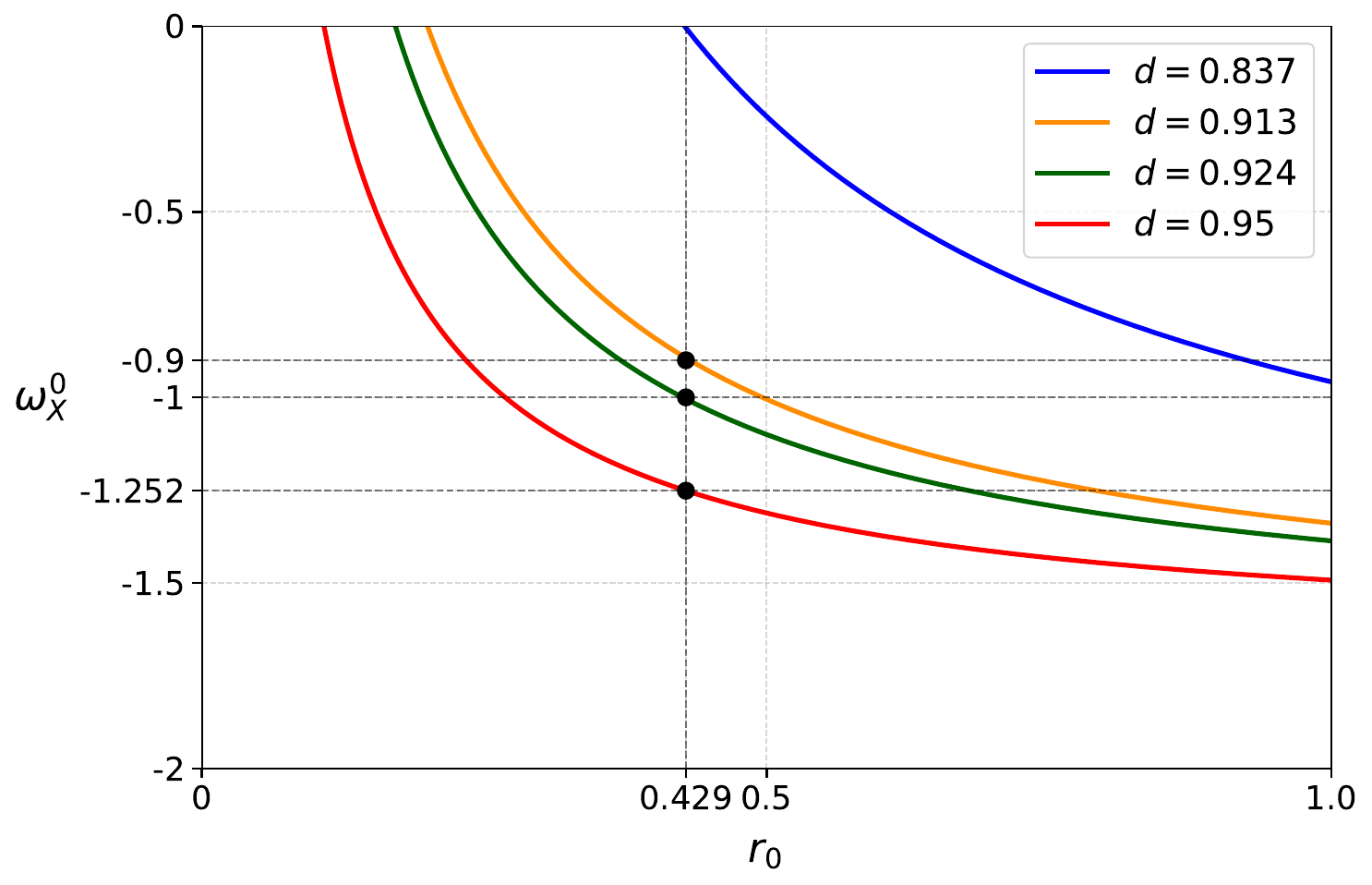}
    \caption{The range of the free parameter, $0.913 \lesssim d \lesssim 0.924$, is compatible with both the current density ratio $r_0 \simeq 0.429$ and the observationally favored range of the current equation of state for dark energy, $-1 \leq \omega_X^0 \lesssim -0.9$.}
    \label{Fig.1}
\end{figure}

In terms of the density ratio, the equation of state \eqref{e.o.s} can be rewritten as
\begin{equation}
    \omega_{X} 
    = \omega_{m}
    - \frac{2}{3}\left(2-q\right) \frac{d^2r+d^2-1}{d^2 r}.
\end{equation}
Here, we assume dust with $\omega_m = 0$.
The observed current values are given by $q_0 \simeq -0.51$ \cite{21}, $\Omega_m^0 \simeq 0.3$, and $\Omega_X^0 \simeq 0.7$ \cite{18}.
The current density ratio is then $r_0 = \rho_m^0 / \rho_X^0 = \Omega_m^0 / \Omega_X^0 \simeq 0.429$.
For $0.837 \lesssim d < 1$, the second term on the right-hand side gives a negative contribution at the current epoch.
Thus, the current equation of state $\omega_X^0$ is shifted toward negative values, opening up the possibility of cosmic acceleration.
As shown in Fig. \ref{Fig.1}, the values of the free parameter that are compatible with both the current density ratio $r_0 \simeq 0.429$ and the observationally favored range $-1 \leq \omega_X^0 \lesssim -0.9$ lie in the range $0.913 \lesssim d \lesssim 0.924$.

We now turn to the dynamical aspect of the cosmic coincidence problem.
Since this problem concerns why the matter density and the dark energy
density are comparable at the present epoch, a key question is whether the
density ratio $r=\rho_m/\rho_X$ can evolve from a larger value in the
matter-dominated era to an order-unity value today.

From Eq.~\eqref{dot r 1}, the sign of $\dot r$ is controlled by the
deceleration parameter $q$. The density ratio decreases when $q<2$.
At the current epoch, this condition is clearly satisfied because the observed
value is $q_0\simeq -0.51$, and hence $\dot r_0<0$.

We next examine whether the same condition is satisfied during the
matter-dominated (MD) era. From Eqs.~\eqref{continuity eq of matter no inter}
and \eqref{torsion scalar scaling behavior}, the matter density and the
torsion scalar scale as $\rho_m\sim a^{-3}$ and $\Phi\sim a^{-3}$,
respectively. Therefore, the torsion contribution appearing in the
Friedmann-like equation scales as $\Phi^2\sim a^{-6}$. Although this term
grows faster than $\rho_m$ toward earlier times, the matter-dominated era is
defined by the regime in which the matter density dominates the effective
Friedmann equation, namely $3M_p^2\Phi^2\ll \rho_m$. In this regime, the first
Friedmann-like equation \eqref{Friedmann-like 1} reduces to
\begin{equation} \label{Friedmann-like 1 MD}
    H^2
    = \frac{1}{3 M_p^2} \left( \rho_m + \rho_X - 3 M_p^2 \Phi^2 \right)
    \simeq \frac{\rho_m}{3 M_p^2}
    \sim a^{-3}.
\end{equation}
This gives the standard matter-dominated behavior $a\sim t^{2/3}$.
Using $q=-a\ddot a/\dot a^2$, one obtains $q_{\rm MD}= 0.5$ at the MD era, which satisfies
$q_{\rm MD}<2$ and hence $\dot r_{\rm MD}<0$.

Consequently, the density ratio decreases both during the matter-dominated
era and at the current epoch. This supports the interpretation that the
current order-unity value of $r$ can arise from the torsion-induced dynamical
evolution of the density ratio, rather than from a constant ratio fixed only
by the free parameter.

At the current epoch, the density ratio \eqref{ratio torsion} is given by
\begin{equation} \label{current ratio}
    r_0
    = \frac{1-d^2}{d^2} 
    + \frac{1}{d^2}\left(\frac{\Phi_0}{H_0}\right)^2.
\end{equation}
In the torsion-free limit, $\Phi_0 \rightarrow 0$, the current density ratio
reduces to $r_0 = (1-d^2)/d^2$. In this limit, an order-unity value of
$r_0$ is not guaranteed over the whole interval $0.837\lesssim d < 1$;
rather, it depends on the choice of the free parameter $d$. In contrast,
a weak but nonzero torsion contribution adds a positive term to $r_0$.

\begin{figure}[h]
    \centering
    \includegraphics[width=\linewidth]{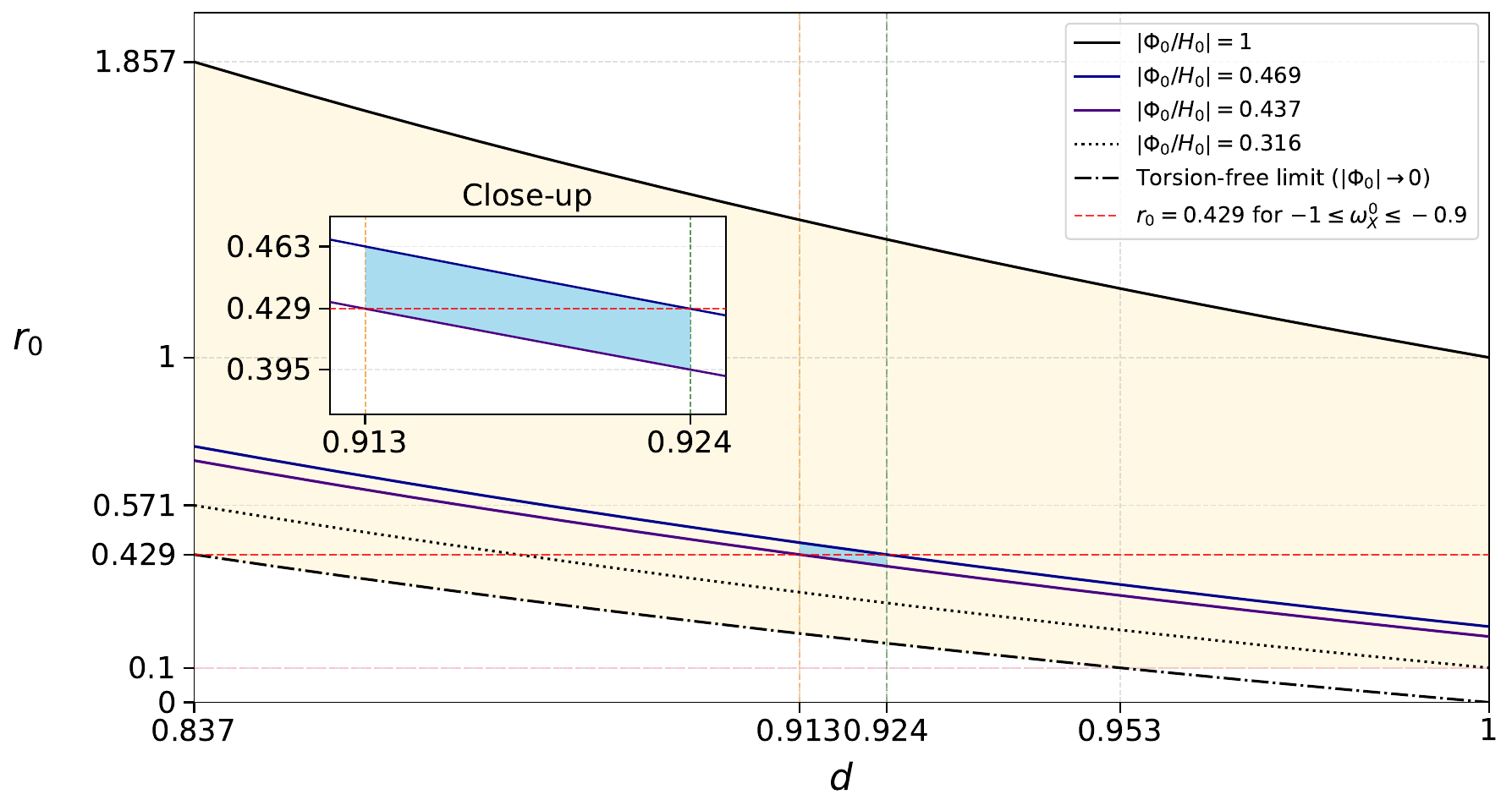}
    \caption{The yellow shaded region denotes the order-unity range of the current density ratio, $r_0=\mathcal{O}(1)$ in the weak torsion regime $|\Phi_0/H_0|<1$, while the red dashed line represents the observationally estimated value $r_0\simeq 0.429$. In the inset, the blue shaded region represents the parameter region bounded by $0.913\lesssim d\lesssim 0.924$ and by the curves corresponding to $|\Phi_0/H_0|=0.437$ and $0.469$. The intersections of the red dashed line with the curves inside this blue region indicate solutions satisfying both $r_0\simeq 0.429$ and $-1\leq \omega_X^0\lesssim -0.9$.}
    \label{Fig.2}
\end{figure}

As shown in Fig. \ref{Fig.2}, for $0.316 \lesssim |\Phi_0/H_0| < 1$
within the weak torsion regime, the current density ratio remains in the
order-unity range throughout the whole interval $0.837\lesssim d < 1$.
Thus, the observed order-unity value of $r_0$ can be interpreted as a
consequence of the torsion-induced contribution, rather than as a result
of tuning or further restricting the free parameter $d$. The observationally estimated value $r_0\simeq 0.429$ can be realized within the weak torsion subregion.

 In the inset of Fig. \ref{Fig.2}, the red dashed line intersects the curves inside the blue shaded region, which is bounded by $0.913\lesssim d\lesssim 0.924$ and by the curves for $|\Phi_0/H_0|=0.437$ and $|\Phi_0/H_0|=0.469$. These intersections indicate solutions that satisfy both $r_0\simeq 0.429$ and the observationally
favored range $-1\leq \omega_X^0\lesssim -0.9$.

\section{Conclusions}

We have investigated a dynamical resolution of the cosmic coincidence problem in non-interacting holographic dark energy with the Hubble radius as the IR cutoff within Einstein-Cartan gravity. We considered the torsion scalar $\Phi$ compatible with the cosmological principle, whose scaling behavior
$\Phi\sim a^{-3}$ is obtained self-consistently. The torsion contribution shifts the equation of state for holographic dark energy away from the dust-like value and toward negative values, allowing cosmic
acceleration without introducing an interaction between dark matter and dark energy. For the current density ratio and an observationally favored range of the equation of state for dark energy, the corresponding holographic free parameter lies in the range $0.913\lesssim d\lesssim 0.924$.

The key result is that the torsion scalar renders the density ratio $r=\rho_m/\rho_X$ time-dependent even in the strictly non-interacting case $Q=0$. We have shown that $\dot r<0$ is satisfied both in the matter-dominated era and at the current epoch, so that the density ratio can decrease dynamically toward its current order-unity value. Moreover, in the weak but nonzero torsion subregion, $0.316\lesssim |\Phi_0/H_0|<1$, the current density ratio can remain in the order-unity range throughout the allowed interval $0.837\lesssim d<1$, without tuning or further restricting the holographic free parameter.

Thus, the coincidence problem is not addressed by a phenomenological dark sector interaction or by tuning the holographic free parameter, but by the torsion-induced dynamical modification of the density ratio. These results suggest that Einstein-Cartan torsion can provide a geometric mechanism
that replaces the phenomenological dark sector interaction and can offer a dynamical resolution of the cosmic coincidence problem while maintaining the possibility of cosmic acceleration.

\section*{Acknowledgements}
This work was supported by National Research Foundation of Korea(NRF) Grants funded by the Korea 
Government(MSIT) (No. NRF-2020R1F1A1068410).

\end{document}